\documentclass{article}
\usepackage{amsmath}
\usepackage{amssymb}
\usepackage{url}
\newtheorem{lemma}{Lemma}
\newtheorem{theorem}{Theorem}
\newtheorem{definition}{Definition}
\newtheorem{claim}{Claim}
\newtheorem{proposition}{Proposition}

\date{}
\title{Partial matching width and its application to lower bounds for branching programs}
\author{Igor Razgon\\Department of Computer Science and Information Systems\\
      Birkbeck University of London\\
    \texttt{igor@dcs.bbk.ac.uk}}
\begin{document}
\maketitle
%Partial matching width
\begin{abstract}
We introduce a new structural graph parameter called \emph{partial matching width}.
In particular, for a graph $G$ and $V \subseteq V(G)$, the matching width of $V$
is such a largest $k$ such that for any permutation $SV$ of vertices with $V$, there is
a prefix $SV'$ of $SV$ such that there is a matching of size $k$ and between $SV'$
(being treated as a set of vertices) and $G \setminus SV'$.

For each (sufficiently large) integer $k \geq 1$, we introduce a class $\mathcal{G}_k$ of graphs of treewidth
at most $k$ and max-degree $7$ such that for each $G \in \mathcal{G}_k$
and each (sufficiently large) $V \subseteq V(G)$, the partial matching width of $V$
is $\Omega(k \log |V|)$.

We use the above lower bound to establish a lower bound on the size of non-deterministic
read-once branching programs (NROBPs). In particular, for each sufficiently large 
ineteger $k$, we introduce a class ${\bf \Phi}_k$ of CNFs of (primal graph) treewidth at
most $k$ such that for any $\varphi \in {\bf \Phi}_k$ and any Boolean function
$F \subseteq \varphi$ and such that $|\varphi|/|F| \leq 2^{\sqrt{n}}$ (here the functions are regarded
as sets of assignments on which they are true), a NROBP implementing $F$ is of size 
$n^{\Omega(k)}$. This result significantly generalises an earlier result of the author
showing a non-FPT lower bound for NROBPs representing CNFs of bounded treewidth. Intuitively,
we show that not only those CNFs but also their arbitrary one side approximations with an exponential
ratio still attain that lower bound.

The non-trivial aspect of this approximation is that due to a small number of satisfying assignments
for $F$, it seems difficult to establish a large bottleneck: the whole function can `sneak' 
through a single rectangle corresponding to just \emph{one} vertex of the purported bottleneck.
We overcome this problem by simultaneously exploring $\sqrt{n}$ bottlenecks and showing
that at least one of them must be large. This approach might be useful for establishing other
lower bounds for branching programs.
\end{abstract}

\section{Introduction}
In this paper we introduce a new structural graph parameter \emph{partial matching width}
defined as follows. Let $G$ be a graph, $V \subseteq V(G)$.
The partial matching width of $V$ is the largest $k$ such 
that for any permutation $SV$ of vertices of $V$, there is
a prefix $SV'$ of $SV$ such that there is a matching of size $k$ and between $SV'$
(being treated as a set of vertices) and $G \setminus SV'$.

The partial matching width generalizes matching width of a graph \cite{RazgonKR},
which is the partial matching width of $V=V(G)$. 
In light of a linear relationship between
matching width and pathwidth \cite{obddtcompsys}, partial matching width can be considered
a generalization of the latter.

We show that, similarly to pathwidth, the partial matching width can be much larger
than the treewidth. In particular, for each (sufficiently large) integer $k \geq 1$, 
we introduce a class $\mathcal{G}_k$ of graphs of treewidth
at most $k$ and max-degree $7$ such that for each $G \in \mathcal{G}_k$
and each (sufficiently large) $V \subseteq V(G)$, the partial matching width of $V$
is $\Omega(k \log |V|)$. This class is essentially the same as we used in \cite{RazgonAlgo}
with the only difference that, for the convenience of the reasoning, instead of the 
underlying binary trees we use ternary ones.

Intuitively, we can say that the partial matching width serves for the above class
as an expansion-like equivalent of pathwidth. A similar in spirit connection 
between treewidth and standard expansion has been established in \cite{GroheM09} 
but in a much more general context. 

We use the above lower bound on partial matching width to prove a lower bound
for read-once branching programs, significantly generalizing our earlier result 
\cite{RazgonAlgo}. In particular, in \cite{RazgonAlgo}, for each sufficiently large $k$ 
we introduced a class ${\bf \Phi}_k$ 
of CNFs whose primal graph is of treewidth at most $k$ and showed that NROBPs 
representing this class must be of size $n^{\Omega(k)}$.
Thus, we demonstrated that NROBPs are not FPT on CNFs of bounded treewidth.
In this paper, we show that this lower bound is very \emph{robust}
because it holds for arbitrary one-side approximations of the functions
of ${\bf \Phi}_k$ with ratio up to $2^{\sqrt{n}}$.
Specifically, we show if we take any function $\varphi \in {\bf \Phi}_k$
and consider an arbitrary function $F$ with $F \subseteq \varphi$
and $|\varphi|/|F| \leq 2^{\sqrt{n}}$ \footnote{In other words, we obtain $F$ 
by arbitrary removal of satisfying assignments of $\varphi$ so that at least 
$2^{-\sqrt{n}}$-th part of the initial satisfying assignment remains.}
the lower bound of $n^{\Omega(k)}$ still holds. \footnote{In this paper we use
slightly tweaked version of the CNFs we used in \cite{RazgonAlgo} replacing the
underlying binary tree in the tree decomposition with ternary one. This makes reasoning
by induction reasoning more elegant but the result itself remains true for both initial
and tweaked classes of CNFs.}

This result has two interesting aspects: the approach that we used and
the connection between the `approximation' lower bound and randomized
branching programs. Let us overview both these aspects.

We overview the approach in comparison with
the one we used in \cite{RazgonAlgo}. In particular, in \cite{RazgonAlgo} we considered
a NROBP $Z$ representing a CNF $\varphi \in {\bf \Phi}_k$ and 
fixed a large \emph{bottleneck} of $Z$: a source-sink cut such that,
for some universal constant $c$, at most $n^{-k/c}$-th path of 
satisfying assignments of $\varphi$ `passes' through a single vertex
of this cut. Then we concluded that the total number of vertices in the
cut must be $n^{\Omega(k)}$. In our approximation case this approach
does not work: if a function $F$ has $|\varphi|/2^{\sqrt{n}}$ satisfying
assignments, \emph{all} of them can `sneak' through a \emph{single} vertex
of the cut! Using \emph{partial} matching width instead of just matching
width allows us to avoid fixing a single bottleneck. Instead, we
\emph{simultaneously} consider $\sqrt{n}$ different bottlenecks
and prove that at least one of them must be large.

This approach may be useful for other cases where the methodology
of establishing a single bottleneck does not seem to work. 
One such notable question
is the complexity of \emph{semantic} NROBPs. For domains of size $r>3$,
exponential lower bounds for $r$-way semantic NROBPs have been known
for long \cite{BeameJS01,Jukna09}, recently culminating with a lower bound for $r=3$
\cite{CookEMP16}. However, to the best of our knowledge, the binary case remains
open. Moreover, \cite{CookEMP16} provides an indication that fixing a single
large bottleneck may not be the right technique for tackling the binary
case. In light of this, it is interesting to investigate whether our
approach of multiple bottlenecks would bring any new insight concerning the
binary case.

Our result is related to Randomized Read-Once branching programs through
a well known result of Sauerhoff \cite{Sauerhoff98} (see also Theorem 11.8.3. of \cite{WegBook})
who showed that a lower bound
for a randomized (deterministic) read-once branching program for a particular
function follows from a deterministic read-once branching program lower
bound for an arbitrary constant approximation of this function.
The approximation of \cite{Sauerhoff98} is different from ours:
it is two sided and taken over the whole set of $2^n$ truth assignments,
not just the satisfying ones. It is interesting to see whether our
approach can yield a lower bound for a two-side approximation for the considered
classes of CNFs (and thus a non-FPT lower bound for randomized read-once branching
programs). A natural initial step is to establish a one sided approximation for
deterministic read-once branching programs representing the negations of 
CNFs ${\bf \Phi}_k$ as above.

%TODO: Further structure of the paper plus possibly mentioning
%additional width parameters. 

The rest of the paper is organized as follows. 
Section \ref{sec:prelim} introduces the necessary background.
Section \ref{sec:pmwidth} introduces the notion of partial matching width
and proves existence of a class of graphs of small treewidth in which each
sufficiently large set of vertices has large partial matching width.
Section \ref{sec:lowerbound} proves a non-FPT lower bound for NROBPs
representing functions approximating CNFs of small primal graph treewidth.
%Finally Section \ref{sec:proportion} proves an auxiliary theorem needed for
%the NROBP lower bound proved in Section \ref{sec:lowerbound}. 
\section{Preliminaries} \label{sec:prelim}

{\bf Sets of literals and variables.}
In this paper when we refer to a  \emph{set of literals} we assume that it does not
contain an occurrence of a variable and its negation.
For a set $S$ of literals we denote by $Var(S)$ the set of variables
whose literals occur in $S$. If $F$ is a Boolean function
or its representation by a specified structure, we denote by $Var(F)$
the set of variables of $F$. A truth assignment to $Var(F)$ on which $F$
is true is called a \emph{satisfying assignment} of $F$. A set $S$ of literals
represents the truth assignment to $Var(S)$ where variables occurring
positively in $S$ (i.e. whose literals in $S$ are positive) are assigned with $true$
and the variables occurring negatively are assigned with $false$.

{\bf Projections, restrictions.}
For $V \subseteq Var(S)$, the \emph{projection} of $S$ on $V$ denoted by
$Proj(S,V)$ is the subset  $S' \subseteq S$ such that $Var(S')=V$.
Let  $F$ be a Boolean function, $S$ be a set of literals such that
$Var(S) \subseteq Var(F)$. The \emph{restriction} $F|_S$ is a function
$Var(F) \setminus Var(S)$ such that $S'$ is a satisfying assignment of 
$F|_S$ if and only if $S \cup S'$ is a satisfying assignment for $F$.
If $S$ consists of a single literal $\ell$ then we write $F|_{\ell}$
rather than $F|_{\{\ell\}}$. 

{\bf Boolean functions as sets of satisfying assignments.}
In this paper we regard Boolean functions and CNFs as their sets of satisfying
assignments. In this context if for instance $F$ and $\varphi$ are CNFs and we
write $F \subseteq \varphi$ this means that each satisfying assignment of $F$ is also 
a satisfying assignment of $\varphi$. We also use $|F|$ and $|\varphi|$ to denote
the sizes of respective sets of satisfying assignments.

{\bf CNFs $\varphi(G)$.}
Let $G$ be a graph without isolated vertices. Then $\varphi(G)$ is a 
CNF with $V$ as the set of variables and $\{(u \vee v)| \{u,v\} \in E(G) \}$
as the set of clauses. This definition allows us to identify variables of
$\varphi(G)$ and vertices of $G$ and to use phrases like
`let $S$ be a set of literals of $V(G)$ and let $V \subseteq V(G)$ be the
set of all $v$ such that $\neg v \in S$'.

\begin{definition}[{\bf Nondeterministic Read-once branching programs (NROBPs).}]
Let $V$ be a set of Boolean variables.
Let $Z$ be a directed acyclic graph (DAG) with one source and one sink so that some
of the edges are labelled with literals of $V$.
We say that $v \in V$ \emph{occurs} on edge $e$ is $Z$ is $Z$ is labelled with a literal
of $v$. The occurrence can be \emph{positive} or \emph{negative} if the labelling literal
is $v$ and $\neg v$, respectively.

We say that $Z$ is a nondeterministic read-once branching program (NROBP) implementing a function
on $V$ if each variable of $V$ occurs \emph{exactly} once \footnote{The requirement of
exactly one occurrence rather than at most one means that the NROBP is uniform. Since it is known that every
NROBP can be simulated by a uniform one with only a polynomial increase of the number of nodes (see e.g. \cite{Okol}),
we assume the uniformity w.l.o.g.} on each source-sink path of $Z$.

For a path $P$ of $Z$ we denote by $A(P)$ the set of literals labelling $P$ and $Var(A(P))$ is denoted by $Var(P)$.
Then the set of satisfying assignments of the function represented by $Z$ consists of all $A(P)$ 
such that $P$ is a source-sink path of $Z$.
\end{definition}

\begin{proposition} \label{samevar}
Let $Z$ be a NROBP and let $P_1,P_2$ be two paths having the same initial
anf final vertices. Then $Var(P_1)=Var(P_2)$.
\end{proposition}

{\bf Separation of sets of variables by a vertex of a NROBP.}
In light of Proposition \ref{samevar}, for NROBP $Z$, a variable $x$, and
a vertex $v$ of $Z$, we can say that $x$ is located \emph{before} $v$
(that is, on each source-sink path $P$ including $v$ $x$ occurs on the prefix
of $P$ ending with $v$) or $x$ is located \emph{after} $v$ (replace the prefix by the suffix).
We say that two sets $X$ and $Y$ of variables are \emph{separated} by $X$ if either
(i) all of $X$ occur before $v$ and all of $Y$ occur after $v$ or (ii) 
all of $Y$ occur before $v$ and all of $X$ occur after $v$.

\begin{proposition} \label{fixedset}
Let $G$ be a graph without isolated vertices,
$F \subseteq \varphi(G)$ and $Z$ be a NROBP representing $F$.
Let $u$ be a vertex of $Z$ and let 
$M=\{\{x_1,y_1\}, \dots, \{x_q,y_q\}\}$ be a matching of $G$ such that
$\{x_1, \dots, x_q\}$ and $\{y_1, \dots, y_q\}$ are separated by $u$.
Then there is a set $X$ of $q$ vertices of $G$ consisting of exactly
one vertex of each $\{x_i,y_i\}$ such that for each source-sink path $P$
passing trough $u$, $X \subseteq A(P)$.
\end{proposition}

{\bf Proof.}
If the statement does not hold then for some $\{x_i,y_i\}$, neither $x_i$
nor $y_i$ belongs to $A(P)$ for each source-sing path passing trough $u$.
It follows that there are two such paths $P_1$ and $P_2$ such that
$\neg x_i \in A(P_1)$ and $\neg y_i \in A(P_2)$.
Assume w.l.o.g. that all of $x_1, \dots, x_q$ occur before $u$ and all of $y_1, \dots, y_q$
occur after $u$. Let $P'_1$ be the prefix of $P_1$ ending with $u$ and let 
$P''_2$ be the suffix of $P_2$ beginning with $u$. Then $P=P'_1+P''_2$ is a source-sink
path pof $Z$ such that $\{\neg x_i,\neg y_i\} \subseteq A(P)$. As $F \subseteq \varphi(G)$
it follows that $A(P)$ is also a satisfying assignment of $\varphi(G)$. However, this is 
a contradiction because $A(P)$ falsifies $(x_i \vee y_i)$.
$\blacksquare$

\section{Partial matching width} \label{sec:pmwidth}
\begin{definition}[{\bf Partial matching width}]
Let $G$ be a graph and let $V \subseteq V(G)$.
The \emph{partial matching width} of $V$ (w.r.t. $G$)
is the largest $k$ such that any permutation $SV$ of $V$
has a prefix $SV'$ such that there is a matching of size at least
$k$ between $SV'$ and the rest of of $G$.
\end{definition}

The main theorem of this section is Theorem \ref{mainptv}, where
we prove that for each sufficiently large $k$ there is a class of graphs with
treewidth at most $k$ such that for each graph $G$ of this class and each
sufficiently large $V \subseteq V(G)$, the partial matching width of 
$V$ is $\Omega(k \log n)$.

For the purpose of proving a lower bound on partial matching width,
it will be easier for us to use a related notion of witnessing matching.

\begin{definition}[{\bf Witnessing matching}]
Let $V \subseteq V(G)$ and let $SV$ be a permutation of $V$.
Let $\{u,v\} \in E(G)$. Then $\{u,v\}$ is \emph{supported}
by a partition $SV_1,SV_2$ into a prefix and a suffix if 
either (i) say, $u \in SV_1$ and $v \in SV_2$ or (ii) say, $u \in SV$, $v \in V(G) \setminus V$.
A matching $M$ is supported by $SV_1,SV_2$ is every edge of $M$ is supported
by $SV_1,SV_2$.

A matching $M$ is \emph{witnessing} for $SV$ if there is a partition of $SV$ into a prefix
$SV_1$ and a suffix $SV_2$ supporting $M$.
\end{definition}

\begin{proposition} \label{witnesswidth}
If every permutation $SV$ of $V \subseteq V(G)$ has a witnessing matching of size at least $k$
then the partial matching width of $V$ is at least $k/2$.
\end{proposition}

{\bf Proof.}
Let $SV$ be a permutation of $V$. We need to show that there is a prefix $SV'$ of $SV$
such that there is a matching of size at least $k/2$ between $SV'$ and $V(G) \setminus SV'$.
Let $M'$ be a witnessing matching of size $k$ for $SV$. 
This means that there is a partition of $SV$ into a prefix $SV_1$ and a suffix $SV_2$
and a partition of $M'$ into $M'_1$ and $M'_2$ such that the edges of $M'_1$
connect vertices of $SV_1$ to vertices of $SV_2$ and edges of $M'_2$ connect vertices
of $SV$ to vertices of $V(G) \setminus V$.

Clearly either $M'_1$ or $M'_2$ is of size $k/2$.
In the former case set $SV'=SV$ and $M=M'_1$, in the latter case set $SV'=SV$ and $M=M'_2$.
Clearly, in both cases, we have a matching of size at least $k/2$ connecting the chosen prefix 
to the rest fo the graph.
$\blacksquare$

%%%%%%%%%%%%%%%%%%%%%%%Reasoning for incomplete sets%%%%%%%%%%%%%%%%%%%
{\bf Graphs $T(H)$.}
Let $T$ be a tree and $H$ be an arbitrary graph. 
The graph $T(H)$ is the union of disjoint copies $H^v$ of $H$ for each
$v \in V(T)$ plus additional edges defined as follows.
Let $V(H)=\{1, \dots, p\}$. For $ \in \{1, \dots, p\}$, denote the
copy of $i$ in $H^v$ by $v^i$. Then for each $i \in \{1, \dots, p\}$
there is an edge bewteen $u^i$ and $v^i$ whenever $u$ and $v$ are adjacent 
in $T$.

Let $V\subseteq T(H)$.
$u \in V(T)$ is \emph{occupied} by $V$ if $V(H^u) \cap V \neq \emptyset$.
$OC(T,V)$ denotes the set of all vertices of $T$ occupied by $V$.
If $V(H^u) \subseteq V$ then we say that vertex $u$ is \emph{complete}
in $V$. 

Suppose $V \subseteq V(T(H))$, and let $V_1,V_2 \subseteq V$ (one of $V_1,V_2$ possibly empty)
such that $V_1 \cup V_2=V$ and $V_1 \cap V_2=\emptyset$. Then the vertices 
of $T(H)$ may have three different \emph{roles} w.r.t. $V_1,V_2$: belonging to $V_1$, 
belonging to $V_2$, belonging to $V(T(H)) \setminus V$.
A vertex $u \in V(T)$ is \emph{homogenous} if all the vertices of
$V(H^u)$ have the same role.

{\bf Ternary trees and the $tr$ function.}
A complete rooted ternary tree of heigh $h$ is a tree with a special designated root vertex $rt$
(naturally determining the parent-child relation between the vertices)
in which each root-leaf path has exactly $h$ edges and each non-leaf vertex has exactly
$3$ children. For $x \geq 1$, we denote by $tr(x)$ the largest $h$ such that
is at least as the number of vertices of a complete rooted ternary tree $T$ of height $h$.
It is not hard to see that $tr$ is a logarithmic function and that
$|V(T)| \leq x \leq 3|V(T)|$.

\begin{lemma} \label{goodpart2}
Let $T$ be a tree and $H$ a graph. Let $V \subseteq V(T(H))$,
$V_1,V_2 \subseteq V$, $V_1 \cup V_2=V$, $V_1 \cap V_2=\emptyset$.
Assume that there are two vertices $u$ and $v$ of $T$
and a subset $\{1, \dots, t\}$ of vertices of $H$ such that
for each $i \in \{1, \dots, t\}$, the role of $u^i$ is not
the same as the role of $v^i$.
Then $T(H)$ has a matching $M$ of size $t$ such the ends
of each edge $e$ of $M$ have different roles.
\end{lemma}

{\bf Proof.}
Let $P=u_1, \dots, u_q$ be the path between $u$ and $v$
such that $u_1=u$ and $u_q=v$.
Then for each $i \in \{1, \dots, t\}$,
$P^i=u_1^i, \dots, u_q^i$ is a path where the first and the 
last vertices have different roles. It follows that $P^i$
has an edge $e^i$ whose ends have different roles.
Each such $e^i$ connects two copies of vertex $i$ of $H$,
hence for $i \neq j$, edges $e^i$ and $e^j$ cannot have a 
joint end. It follows that edges $e^1, \dots, e^t$
constitute a matching of size $t$.
$\blacksquare$

\begin{lemma} \label{goodpart3}
Let $T$ be a tree with at least $p$ vertices.
Let $H$ be a connected graph of at least $2p$ vertices.
Let $V \subseteq V(T(H))$ such that $OC(T,V)=V(T)$. 
Let $V_1,V_2$ be a partiton of $V$ so that 
$|V_i| \leq |V(T(H))|-p^2$ for each $i \in \{1,2\}$. 
Then $T(H)$ has a matching of size $p$ in which
the ends of each edge have different roles w.r.t. $V_1,V_2$. 
\end{lemma}

{\bf Proof.}
Assume first that $T$ has at least $p$ non-homogenous vertices.
%vertices $u$ such that
%not all vertices of $H^u$ have the same role. 
Then, due to 
the connectedness of $H$, for each non-homogenous vertex $u$, 
there is an edge $e_u$ of $H^u$
whose ends have different roles.
As these edges belong to different copies of $H$, no two of them have
a joint end and hence they constitute a matching of size $p$.
%Arguing as in the first part of Lemma \ref{goodpart1},
%these edges form a desired matching of size at least $p$.

%In order to continue, we need an auxiliary claim.

If $T$ does not have $p$ non-homogenous vertices, it has 
at least one homogenous vertex $u$. 
As by assumption $V(H^u)$ intersects
with $V$, it is either that $V(H^u) \subseteq V_1$
or $V(H^u) \subseteq V_2$. W.l.o.g., assume the former.
If $T$ has another vertex $v$ such that $V(H^v) \subseteq V_2$
then we are immediately done by Lemma \ref{goodpart2}.
Thus, we conclude that for all vertices $v$ of $V(T)$
except at most $p-1$ ones, $(H^v) \subseteq V_1$.

Observe that there is $w \in V(T)$ such that 
$|V(H^w) \setminus V_1| \geq p$. Indeed, assume the
opposite. Let $W$ be the set of most $p-1$ 
non-homogenous vertices. 
It follows that each $w \in W$ has at most $p-1$ vertices
outside $V_1$ and hence, the total number of vertices 
that are not in $V_1$ is at most $(p-1)^2<p^2$
in contradiction to our assumption.
Thus the desired vertex $w$ does exist.

Denote by $\{1, \dots, p\}$ the vertices of $H$ such that
$w^i \notin V_1$ for each $i \in \{1, \dots, p\}$.
As $V(H^u) \subseteq V_1$, $u^i \in V_1$ for all $i \in \{1, \dots, p\}$.
Then the desired matching exists by Lemma \ref{goodpart2}
as witnessed by vertices $u,w$,and $\{1, \dots, p\}$.
$\blacksquare$

\begin{lemma} \label{perfpart}
Let $p \geq 1$ be a natural number, $H$ be a connected graph with at least $2p$ vertices
and $T$ be a complete ternary tree of at least $p$ vertices.
Let $V \subseteq V(T(H))$ with $OC(T,V)=V(T)$ 
and let $SV$ be a permutation of $SV$.
Then $T(H)$ has a witnessing matching $M$ for $SV$ of size at least
$p(tr(|T|)-tr(p))$. Moreover, $M$ is supported by a partition 
$SV_1,SV_2$ of $SV$ into a prefix and a suffix that is \emph{balanced}
in the following sense. For each $i \in \{1,2\}$
$|V(T(H))|-|SV_i| \geq p^2$.
\end{lemma}

{\bf Proof.}
By induction on $tr(|T|)$.
Assume first that $tr(|T|)-tr(p) \leq 1$.
Partition $SV$ into a prefix $SV_1$ and a suffix $SV_2$
of size different by at most $1$. Note that this partition
is balanced. Indeed, if $|SV| \geq 2p^2$ then $|SV_i| \geq p^2$
for each $i \in \{1,2\}$. Then $|V(T(H))|-|SV_i| \geq
|SV|-|SV_i|=|SV_{3-i}| \geq p^2$.
Otherwise, $|SV_i| \leq p^2$ for each $i \in \{1,2\}$
Note also that $|V(T(H))| \geq 2p^2$: at least $p$ vertices
each associated with a copy of $H$ of at least $2p$.
Hence $|V(T(H))|-|SV_i| \geq p^2$.
It follows from Lemma \ref{goodpart3} that $T(H)$ has a matching $M$
of size at least $p$ with the ends of each edge having different
roles w.r.t. $SV_1,SV_2$ seen as sets of vertices. Clearly,
$M$ is a witnessing matching for $SV$ supported by $SV_1,SV_2$.
Hence, the lemma holds in the considered case. 

Assume now that $tr(|T|)-tr(p) \geq 2$.
Let $rt$ be the root of $T$ and let $T_1,T_2,T_3$ be the subtrees
of $T$ rooted by the children of $T$. For $i \in \{1,2,3\}$
let $V_i=V \cap V(T_i(H))$ and $SV_i$ be the permutation of $V_i$
where the order of elements is the same as in $SV$. Note that as the height of $T_i$ is one less than
that of $T$, $tr(|T_i|) \geq tr(p)+1$ and hence $|T_i| \geq p$, hence
the lemma is correct for $T_i,V_i,SV_i$ by the induction assumption.
It follows that $T_i(H)$ has a witnessing matching $M_i$ for
$SV_i$ having size at least  $p(tr(|T|)-tr(p))-p$ and supported
by a balanced partition $SV'_i,SV''_i$ of $SV_i$ into a prefix and 
a suffix. 

Let $u_1,u_2,u_3$ be the final vertices of $SV'_1,SV'_2,SV'_3$,
respectively. Assume w.l.o.g. that they occur in 
$SV$ in the order they are listed. Let $SV_1$ be the prefix of $SV$
ending with $u_2$. Let $SV_2=SV \setminus SV_1$.

Let $T^*=T \setminus T_2$, $V^*=V \cap V(T^*(H))$.
For $i \in \{1,2\}$, let $V^*_i=V^* \cap SV_i$ (the latter is treated
as a set). Then for each $i \in \{1,2\}$
$V(T^*(H)) \setminus V^*_i \geq	p^2$. 
Indeed, note that $SV'_1=V^*_1 \cap V(T_1(H))$ and 
$SV''_3=V^*_2 \cap V(T_2(H))$. 
Therefore, $|V(T^*(H)) \setminus V^*_1| \geq
|V(T_1(H)) \setminus V^*_1|=|V(T_1(H)) \setminus SV'_1| \geq p^2$,
the last inequality follows from the induction assumption.
Symmetrically 
$|V(T^*(H)) \setminus V^*_2| \geq
|V(T_3(H)) \setminus V^*_2|=|V(T_3(H)) \setminus SV''_3| \geq p^2$.
It follows from Lemma \ref{goodpart3} that $T^*(H)$ has a matching $M^*$
of size $p$ where the ends of each edge have different roles w.r.t.
$V^*_1$ and $V^*_2$. 

Let $M=M_2 \cup M^*$. As $M_2$ and $M^*$ are matchings in vertex-disjoint
subgraphs of $T(H)$, $M$ is a matching of size $|M_2|+|M^*| \geq p(tr(|T|)-tr(p))$.
We claim that $M$ is in fact a witnessing matching for $SV$ supported by $SV_1,SV_2$.
Indeed, let $\{u,v\} \in M$. If $\{u,v\} \in M^*$ then either, say $u \in V^* \subseteq SV_1$
and $v^* \in V^*_2 \subseteq SV_2$ or, say $u \in V^* \subseteq SV$ and $v \in V(T^*(H)) \setminus V^*=
V(T^*(H)) \setminus SV \subseteq V(T(H)) \setminus SV$. That is, $\{u,v\}$ is supported by $SV_1,SV_2$ in this
case. It remains to assume that $\{u,v\} \in M_2$. Then 
either, say $u \in SV'_2 \subseteq SV_1$ and $v \in SV''_2 \subseteq SV_2$ or, say
$u \in SV_2 \subseteq SV$ and $v \in V(T_2(H)) \setminus SV_2=V(T_2(H)) \setminus SV \subseteq V(T(H)) \setminus SV$,
confirming again that $\{u,v\}$ is supported by $SV_,SV_2$.

It remains to verify that $SV_1$ and $SV_2$ satisfy the balancing constraints.
For that, notice that for each $i \in \{1,2\}$, 
$T^*(H) \setminus V^*_i \subseteq T^*(H) \setminus SV_i \subseteq T(H) \setminus SV_i$,
as we have already proved that $|T^*(H) \setminus V_i| \geq p^2$, the balancing constraints
follow.
$\blacksquare$

{\bf Immediate subtrees.}
Let $T$ be a complete rooted binary tree.
Then $T'$ is an \emph{immediate} subtree of $T$ if $T'$ is rooted
by a child of the root of $T$. Let $V \subseteq V(T(H))$.
Then $T'$ is the \emph{largest} immediate subtree w.r.t. $V$ if
$|OC(T,V) \cap V(T')|$ is the largest among the immediate subtrees
of $T$.

\begin{definition}
Let $H$ be a graph.
Let $T$ be a complete rooted ternary tree and let $V \subseteq V(T(H))$.
Let us sequences $T_1, \dots, T_q$ and $V_1, \dots, V_q$ as
follows.
\begin{itemize}
\item $T_1=T$, $V_1=V$.
\item Assume that for $1 \leq i<q$, $T_i$ and $V_i$ have been 
defined. Then $T_{i+1}$ is an immediate largest subtree of
$T_i$ w.r.t. $V_i$ and $V_{i+1}=V_i \cap V(T_{i+1}(H))$.
\end{itemize}

Then $T_1, \dots, T_q$ is called a \emph{sequence of largest 
subtrees} of $T$ w.r.t. $V$ and $V_1, \dots, V_q$ is the respective
\emph{sequence of sets}. 

Assume that $|OC(T_1,V_1)|-|OC(T_q,V_q)| \geq p$, while $|OC(T_1,V_1)|-|OC(T_{q-1},V_{q-1})|<p$.
Then we say that $T_1, \dots, T_q$ is a 
\emph{minimal sequence of largest subtrees} of $T$ w.r.t. $V$  \emph{lacking} $p$.
\end{definition}

%It is important to note that for the definition to be well-formed
%it is necessary that each $T_1, \dots, T_{q-1}$ is a complete ternary
%rooted tree.

\begin{lemma} \label{immed2}
Assume that $OC(T,V)>p$.
Then there exists a minimal sequence of largest subtrees of $T$ w.r.t. $V$
lacking $p$.
\end{lemma}

{\bf Proof.}
Let $T_1, \dots, T_q$ be a sequence of largest subtrees of $T$ w.r.t. $V$
such that $|V(T_q)|=1$ 
and $V_1, \dots, V_q$ be the corresponding sequence of sets
(such a sequence clearly exists: if the height of $T$ is $n-1$
then a sequence of largest immediate trees of $n$ elements will be one).
Then $|OC(T_1,V_1)|>p$ and $|OC(T_q,V_q)| \leq 1$. Then, take a minimal subsequence
$T_1, \dots, T_{q'}$ of $T_1, \dots, T_q$ such that
$|OC(T_1,V_1)|-|OC(T_{q'},V_{q'})| \geq p$.
Clearly, $T_1, \dots, T_{q'}$ is a desired sequence.
$\blacksquare$

\begin{lemma} \label{immed3}
Let  $T_1, \dots, T_q$ be a 
\emph{minimal sequence of largest subtrees} of $T$ w.r.t. $V$  \emph{lacking} $p$
and let $V_1, \dots, V_q$ be the corresponding sequence of sets.
Then $|OC(T_q,V_q)| \geq (|OC(T_1,V_1)|-p)/3$.
\end{lemma}

{\bf Proof.}
\begin{claim}
Let $T'$ be a complete rooted ternary tree of height at least $1$ and let
$T'_1,T'_2,T'_3$ be immediate subtrees of $T'$.
Let $V' \subseteq V(T'(H))$ and let $V'_i=V' \cap V(T'_i(H))$.
Let $i \in \{1,2,3\}$ be such that $OC(T'_i,V'_i)$ is the largest.
Then $|OC(T'_i,V'_i)| \geq (|OC(T',V')|-1)/3$. 
\end{claim}

{\bf Proof.}
It is not hard to see that 
$|OC(T',V')|=|OC(T',V') \cap V(T_1)|+|OC(T',V') \cap V(T_2)|+|OC(T',V') \cap V(T_3)|+1$
Further on, it is not hard to see that 
for each $j \in \{1,2,3\}$, $OC(T'_j,V'_j)=OC(T',V') \cap V(T'_j)$.
That is,
$|OC(T',V')|=|OC(T'_1,V'_1)|+|OC(T'_2,V'_2)|+|OC(T'_3,V'_3)|+1$.
Clearly, the largest of the set sizes on the right hand side is at least
one third of $|OC(T',V')|-1$.
$\square$  

By minimality of $q$,
$|OC(T_{q-1},V_{q-1})| \geq |OC(T_1,V_1)|-(p-1)$.
By the above claim,
$|OC(T_q,V_q)| \geq (|OC(T_{q-1},V_{q-1})|-1)/3 \geq (|OC(T_1,V_1)|-(p-1)-1)/3$,
so the desired inequality follows.
$\blacksquare$

\begin{lemma} \label{immed4}
Let  $T_1, \dots, T_q$ be a 
\emph{minimal sequence of largest subtrees} of $T$ w.r.t. $V$  \emph{lacking} $p$
and let $V_1, \dots, V_q$ be the corresponding sequence of sets.
Suppose that $OC(T,V) \subset V(T)$.
Let $T^*=T \setminus T_q$ and let $V^*=V \cap V(T^*(H))$.
Then $OC(T^*,V^*) \subset V(T^*)$.
\end{lemma}

{\bf Proof.}
Let $T'=T \setminus T_2$ and let $V'=V \cap T'(H)$.
\begin{claim}
$OC(T',V') \subset V(T')$.
\end{claim}

{\bf Proof.}
Let $rt$ be the root of $T$.
By definition, $T_2$ is the subtree of $T$ whose root is one of children of $rt$.
If we assume that $OC(T',V')=V(T')$ then both $V(H^{rt})$ and $V(H^u)$ for each $u \in T'$
have non-empty intersections with $V$. In particular, for any ohter child $T'_2$ of $T$,
each copy of $H$ of \emph{each} vertex of $T'_2$ has a non-empty intersection with $V$.
By selection, $T_2$ has the \emph{largest} number of vertices whose copies of $H$ intersect
with $V$. Then, for this maximality to be true, the copy of $H$ associated with \emph{each}
vertex of $T_2$ must have a non-empty intersection with $V$ too.
But this means that the copes of $H$ of \emph{all} the vertices of $T$ have a non-empty intersection
with $V$ in contradcition to our assumption that $OC(T,V) \subset V(T)$.
$\square$

If $q=2$ then we are done by the above claim.
Otherwise, let $u \in V(T')$ such that $V(H^u) \cap V'=\emptyset$.
Note that $V^*$ is obtained by adding to $V'$ the elements of $V$ intersecting
$V(H^v)$ for $v \in V(T^*) \setminus V(T')$.
This means that the intersection of $V^*$ with $H(T^u)$
remains empty.
$\blacksquare$ 

\begin{lemma} \label{goodpart1}
Suppose that $H$ is a connected graph of at least $p$ vertices.
Assume that $|OC(T,V)| \geq p$ and that $V(T) \setminus OC(T,V) \neq \emptyset$. 
Then $T(H)$ has a matching of size $p$ all edges of which have one end in 
$V$ and the other end outside $V$.
\end{lemma}

{\bf Proof.}
Assume first that all the vertices of $OC(T,V)$ are incomplete. 
Then, as $H$ is connected, for each $u \in OC(T,V)$, there is an
edge $e_u$ connecting a vertex of $V(H^u) \cap V$ with a vetrex
$V(H^u) \setminus V$. Let $M$ be the set of these edges.
As they belong to different copies of $H$, they cannot have joint ends and hence
$M$ is a matching. As  $|OC(T,V)| \geq p$ and $|M|=|OC(T,V)|$, $M$ is a matching
required by the lemma.

Otherwise, let $u \in OC(T,V)$ be a complete vertex and let $v$ be a vertex
such such that $V(H^v) \cap V=\emptyset$.
Then the statement immediately follows from lemma \ref{goodpart2}
by taking $V_1=V$ and $V_2=\emptyset$.
$\blacksquare$

\begin{theorem} \label{mwmain}
Let $T$ be a complete rooted ternary tree.
Let $H$ be a connected graph of at least $2p$ vertices.
Let $V \subseteq V(T(H))$ such that $|OC(T,V)| \geq p$.
Let $x=tr(|OC(T,V)|)$. 
Finally, let $SV$ be a permutation of $V$.
Then $T(H)$ has a witnessing matching for $SV$ of size
at least $p*\lfloor(x-tr(p))/2 \rfloor$.
\end{theorem}

{\bf Proof.}
By induction on $|OC(T,V)|$.
As a matching size is non-negative,
the statement is trivially true if $x-tr(p)<2$. 
Assume now that $x-tr(p) \geq 2$.
Then, clearly, $|OC(T,V)|>p$.
We claim that in this case there is a witnessing matching for $SV$ of size at least
$p$. Indeed, if $OC(T,V)=V(T)$, this follows from 
Lemma \ref{perfpart}, otherwise, this follows from Lemma \ref{goodpart1}.
This establishes the statement of theorem for the case where $x-p<4$.
Assume now that $x-tr(p) \geq 4$ and that the theorem has been established
for all the smaller values of $OC(T,V)$.

If $OC(T,V)=V(T)$ then the statement follows from Lemma \ref{perfpart},
hence we assume that $OC(T,V) \subset V(T)$.

Let $T_1, \dots, T_q$ be a
\emph{minimal sequence of largest subtrees} of $T$ w.r.t. $V$  \emph{lacking} $p$
existing by Lemma \ref{immed2}. Let $V_1, \dots, V_q$ be the corresponding sequence of sets.
By Lemma \ref{immed3}, $|OC(T_q,V_q)| \geq (|OC(T,V)|-p)/3$.
Our next step is to apply the induction assumption to $T_q$ and $V_q$.
In order to do this, we must verify that (i) $|OC(T_q,V_q)|<|OC(T,V)|$
and (ii) $|OC(T_q,V_q)| \geq p$. Now, (i) follows by construction.
To show (ii), let us perform the following calculation.

%Let $y=tr(|OC(T_q,V_q)|)$. 
%By Lemma \ref{goodmeasure},
%$y \geq x-2$ and hence $y \geq p+2$ and hence  
%$|OC(T_q,V_q)| \geq p$.
%On the other hand, by construction $|OC(T_q,V_q)|<|OC(T,V)|$ and hence
%inudction can be applied regarding $T_q$ and $V_q$.

Let $T'$ and $T''$ be complete rooted ternary trees of height $tr(p)+1$
and $tr(|OC(T,V)|)$, respectively. Then $|T'|>p$ and $|T''| \leq |OC(T,V)|$.
The assumption $tr(|OC(T,V)|)-tr(p) \geq 4$ implies that the height of
$|T''|$ is greater than the height of $T'$ by at least $3$.
It follows that $|OC(T,V)|\geq  |T''| \geq 27|T'| \geq 27p$ implying (ii).

Let $y=tr(|OC(T_q,V_q)|)$. Let $SV_q$ be the permutation of $V_q$ where the order of elements is the same as in $SV$. 
By the induction assumption, $T_q(H)$ has a witnessing matching $M_q$
for $SV_q$ having size at least $p*\lfloor(y-tr(p))/2 \rfloor$.

Let $T'=T \setminus T_q$ and let $V'=V \cap V(T'(H))$.
By Lemma \ref{immed4}, $OC(T',V') \subset V(T')$.
Taking into account that $OC(T',V') \geq p$ by construction, it follows from 
Lemma \ref{goodpart1} that $T'(H)$ has a matching $M'$ of size at least
$p$ between $V'$ and $T'(H) \setminus V'$.

As $M'$ and $M_q$ are matching in vertex-disjoint subgraphs of $T(H)$,
$M=M' \cup M_q$ is also a matching of size $|M'|+|M_q|$.
We claim that $M$ is in fact a witnessing matching for $SV$ of size at
least $p*\lfloor(x-tr(p))/2 \rfloor$ thus implying the theorem.

As $M_q$ is a witnessing matching of $T_q(H)$ for $SV_q$,
$SV_q$ has a prefix $SV'_q$ such that for any edge $\{u,v\}$
either $u \in SV'_q$ and $v \in SV_q \setminus SV'_q$ or
(ii) $u \in SV_q$ and $v \in TV_q \setminus SV_q$.
Let $w$ be the last vertex of $SV'_q$ and let $SV'$ be the prefix
of $SV$ ending with $w$.

Now, let $\{u,v\} \in M$. Assume first that $\{u,v\} \in M_q$.
If $\{u,v\}$ satisfies condition (i) in  the previous
paragraph then it is not hard to see that $u \in SV'$ and $v \in SV \setminus SV'$
(because by construction $SV'_q \subseteq SV'$ and $SV_q \setminus SV'_q \subseteq SV \setminus SV'$).
If $\{u,v\}$ is of type (ii) then, clearly $u \in SV$ and, as $SV \setminus SV_q \subseteq V(T')$,
$v$ does not belong to $SV$.
Assume now that $\{u,v\} \in M'$. Then, say, $u \in V'$ and hence $u \in SV$.
Also, $v \in T'(H) \setminus V'$ and hence $v \in T(H) \setminus V$ as $V \setminus V'$ is 
a subset of $V(T_q(H))$ disjoint with $V(T'(H))$. Thus we have established $M$ is a witnessing
matching for $SV$ where $SV'$ serves as a witnessing prefix.

It remains to verify that $M$ is of a required size. For this, let us first observe that $y \geq x-2$.
Indeed, let $T'$ and $T''$ be complete rooted ternary trees of height $x$ and $x-2$ respectively.
Note that by definition of $x$, $|T'| \leq |OC(T,V)|$ and, by definition of $y$, it is sufficient
to show that $|T''| \leq |OC(T_q,V_q)|$.

Then,
\begin{equation} \label{eq28}
|T''|=\frac{|T'|-4}{9} \leq \frac{|OC(T,V)|-4}{9} \leq \frac{3|OC(T_q,V_q)|+p-4}{9} \leq 
\frac{|OC(T_q,V_q)|}{3}+\frac{p-4}{9}
\end{equation}
where the second equality follow $|OC(T_q,V_q)| \geq (|OC(T,V)|-p)/3$ proven earlier.
Clearly, $(p-4)/9 \leq 2|OC(T_q,V_q)|/3$ will immediately imply $|T''| \leq |OC(T_q,V_q)|$.
Assume the opposite. Then as $|OC(T_q,V_q)| \geq p$ this means that $(p-4)/9>2p/3$ that
is $5p+4<0$, a contradiction due to the non-negativity of $p$. Thus $y \geq x-2$ has
been established. 

Now, $|M|=|M_q|+|M'| \geq p*\lfloor(x-2-tr(p))/2 \rfloor+p=p*\lfloor (x-tr(p)/2)-1 \rfloor+p=
p*\lfloor (x-tr(p)/2)\rfloor$, as required.
$\blacksquare$

\begin{theorem} \label{mainptv}
There are constants $c_0,c_1,c_2 \geq 1$ such that the following is true.
There is an infinite set of integer numbers $k \geq c_0$
such that for each $k$ there is a class $\mathcal{G}_k$ of treewidth at most $k$
and such that for any $G \in \mathcal{G}_k$ and for any $V \subseteq V(G)$ of
size at least $k^{c_1}$, the partial matching width of $V$ in $G$ is at least
$(k \log |V|)/c_2$.
\end{theorem}

{\bf Proof.}
Consider first numbers $k>0$ that are multiples of $4$.
For each such $k$, let $\mathcal{G}_k$ be the set 
of all graphs $T(H)$ where $H$ is a path of $k/2$ vertices
and $T$ is a complete rooted ternary tree. 
It is not hard to see that the treewidth of each such a graph is at most
$k$: let $T$ be the underlying tree decomposition, the bag of the root
vertex include the root copy of $H$, and the bag of each non-root
vertex include its own copy of $H$ plus that of the parent.

Put $p=k/4$. Then by Theorem \ref{mwmain},
for each $V \subseteq T(H)$ and each permutation
$SV$ of $V$ there is a witnessing matching for $SV$ of size
$p*\lfloor(|OC(T,V)|-tr(p))/2 \rfloor \geq p*\lfloor(tr(|V|/2p)-tr(p))/2 \rfloor$.
As for a sufficiently large $x$, $\log x/2 \leq tr(x) \leq \log x$,
it is not hard to observe that there are constants 
$d_0,d_1,d_2$ such that for each $p \geq d_0$ and $|V| \geq p^{d_1}$,
$p*\lfloor(tr(|V|/2p)-tr(p))/2 \rfloor \geq (p \log |V|)/d_2=(k\log V)/4d_2$.
By Proposition \ref{witnesswidth}, the partial matching width of
$V$ is at least $(k\log V)/4d_2$. Now, let $c_0=4d_0$, $c_1=d_1$ and $c_2=8d_2$.

For numbers $k$ that are not necessarily multiples of $k$, set $c_2:=2*c_2$
and $p=\lfloor k/4 \rfloor$.
$\blacksquare$

\section{A branching program lower bound involving partial matching width} \label{sec:lowerbound}
In this section we prove the following theorem.

\begin{theorem} \label{mainbound}
For each sufficiently large $k$, there is a class ${\bf \Phi}_k$ of CNFs of primal
treewidth at most $k$ such that for each $\varphi \in {\bf \Phi}_k$ and each
$F \subseteq \varphi$ such that $|\varphi|/|F| \leq 2^{\sqrt{n}}$, a NROBP representing
$F$ is of size $n^{\log k/c}$ for some universal constant $c$. 
\end{theorem}

Let $c_0,c_1$ be constants as in Theorem \ref{mainptv}.
Then ${\bf \Phi}_k=\{\varphi(G)|G \in \mathcal{G}_k, |V(G)| \geq k^{2c_1}\}$.
We introduce the lower bound on the number of vertices of $G$ so as to make sure
that the matching width lower bound as specified in Theorem \ref{mainptv}
holds for any $V \subseteq V(G)$ st. $|V| \geq \sqrt{n}$.

An important property of the CNFs $\varphi(G)$ is that, as a result
of fixing many positive literals, the number of satisfying assignments
decreases exponentially.  

To make the above statement more precise, we need to instroduce additional
notation. Let $F'$ be a Boolean function, let $S'$ be a set of literals
with $Var(S') \subseteq Var(F')$. Then $F'\leftarrow S'$ denotes the Boolean
function with the set of satisfying assignments $\{S|S \in F', S' \subseteq S\}$.
That is, $F' \leftarrow S'$ consists of those satisfying assigments of $F'$
that include $S'$ as a subset. Alternatively, $F' \leftarrow S'$ can be obtained
by adding $S'$ to each satisfying assignment of $F'|_{S'}$.

\begin{theorem} \label{manyvars1}
For each $d$ there is a constant $b_d>1$ such that the following is true.
Let $G$ be a graph of max-degree $d$ and let $U \subseteq V(G)$.
Then $|\varphi(G)\leftarrow U| \leq |\varphi(G)|/ 2^{|U|/b_d}$.
\end{theorem}

The proof of Theorem \ref{manyvars1} is provided in the appendix.

%This theorem is proved using the solutions counting decision tree
%in the same way as I did this for my IPEC paper. The main point is that
%the wieght on the edges are constant, concrete constants do not matter.
%I will leave the statement as a stub for the moment, will reproduce
%the proof when the result is ready.

%Let $H$ be a path of $2p$ vertices. 
%Let $TR$ be a complete ternary tree of height $x$.
%For the sake of convenience, denote $\varphi(TR(H))$ by $\varphi(TR,H)$ 
%For convenience, we will denote the latter by $F(TR,H)$.
%Denote $|V(TR(H))|$ by $n$.

The proof of Theorem \ref{mainbound}, is based on simultaneous exploration
of $\sqrt{n}$ bottlenecks. In order to highlight the nee for multipe bottlnecks,
we first prove Theorem \ref{restrbound} below using only a single bottlneck.
Theorem \ref{restrbound} is a restricted version of Theorem \ref{mainbound}
in which the approximation ratio is bounded by a constant. Then we show
why this approach does not seem to work when the apprxomation ratio is
bounded by $2^{\sqrt{n}}$ and provide anactual proof for Theorem \ref{mainbound}.

\begin{theorem} \label{restrbound}
Let $\varphi \in {\bf \Phi}_k$ ad let $F \subseteq \varphi$ be such
that $|\varphi|/|F| \leq 2$
Let $Z$ be a NROBP solving $F$.
Then the size of $Z$ is $n^{\Omega(k)}$.
\end{theorem}

{\bf Proof.}
%It is known that we can consider $Z$ to be uniform.
Let $P$ be a source-sink path of $Z$.
The variable occurrences on $P$ form a permutation $SV$ of $V(G)$.
It follows from Theorem \ref{mainptv} that there is  a prefix
$SV'$ of $SV$ such that there is a matching $M$ between
$SV'$ and $V(G) \setminus SV'=SV \setminus SV'$ of size $k \log n/c_2$, where 
$c_2$ is the constant as in Theorem \ref{mainptv}.

Let $P_1$ be the prefix of $P$ such that $var(P_1)=SV'$ and let $P_2$
be the remaining suffix of $P$. Let $a=a(P)$ be the final vertex of $P_1$ (and the
initial vertex of $P_2)$. Clearly, the ends of each edge of $M$ are separated by $a$.
Therefore, by Proposition \ref{fixedset}, there is a set of vertices $U_a$ one per edge
of $M$ such that for each path $Q$ passing through $a$, $A(Q) \subseteq F \leftarrow U_a$.

Let $X=\{a_1, \dots, a_q\}$ be the set of vertices $a(P)$ over all source-sink
paths of $Z$. Then we claim that $q=n^{\Omega(k)}$ implying the lower bound. %again better to provide a concrete reference
Indeed, as vertices of $X$ form a source-sink cut, each satisfying assignment of $X$ is carried
through one of these vertices and hence belongs to some $F \leftarrow U_a$.
That is, $F=\bigcup_{a \in X} F \leftarrow U_a$, 
and hence $|F| \leq \sum_{a \in X} |F \leftarrow U_a|$.
Let $a \in X$ such that $|F \leftarrow U_a|$ is the largest one. Then
$|F| \leq q*|F \leftarrow U_a|$ and, since $F \leftarrow U_a \subseteq \varphi \leftarrow U_a$, we conclude
that $|F| \leq q* |\varphi \leftarrow U_a|$.
Then, according to Theorem \ref{manyvars1},
$|F| \leq q*|\varphi|/2^{|U_a|/b_7} \leq q*|\varphi|/2^{k \log n/(b_7*c_2)}=q|\varphi|/n^{k/(b_7*c_2)}$, 
where $b_7$ is as in Theorem \ref{manyvars1}.

On the other hand, by our assumption, $|F| \geq |\varphi|/2$. Combining this with the previous paragraph,
we observe $q|\varphi|/n^{k/(b_7*c_2)} \geq |\varphi|/2$. Hence  $q/n^{k/(b_7*c_2)} \geq 1/2$
from where the desired lower bound on $q$ immediately follows. 
 $\blacksquare$

The above approach works only if $|F|$ is sufficiently large compared with 
$|\varphi|/n^k$: say $|F|=|\varphi|/2$ or $|F|=|\varphi|/n^{\sqrt{k}}$.
If, however, $|F|=|\varphi|/2^{\sqrt{n}}$, all the satsfying assignments of $F$
can go through a single vertex $a(P)$.
Hence, the approach will yield only a trivial lower bound of 1.

We overcome this difficulty by associating each source-sink path of a NROBP representing
$F$ with a tuple rather than with a single vertex as specified below.

\begin{lemma} \label{bigmatchings}
Let $\varphi \in {\bf \Phi}_k$, let $G$ be such that $\varphi=\varphi(G)$  and 
let $F \subseteq \varphi$ .

Let $Z$ be a NROBP representing $F$ and let $P$ be a source-sink
path of $Z$. Then $P$ has vertices $a_1, \dots a_q$,  
($q \leq \Theta(\sqrt{n})$ for some constant $q=\sqrt{n}$ in case $\sqrt{n}$ is integer)
such that there is a set  $U \subseteq V(G)$, $|U|=\Omega(k \log n)$ 
such that for each source-sink path $Q$ passing through all of $a_1, \dots, a_q$,
$U \subseteq A(Q)$.
\end{lemma}

{\bf Proof.}
Throughout the proof, we assume, for the sake of simplicity, that
$\sqrt{n}$ is an integer. At the end of the proof we will briefly
outline a way to adjust the construction to the general case.

Let $q=\sqrt{n}$.
Partition $P$ into subpaths $P_1, \dots, P_q$ (meaning that the first subpath 
start with the source of $Z$, the last subpath ends with the sink and the last vertex
of $P_i$ is the first vertex of $P_{i+1}$) so that $|Var(P_i)|=\sqrt{n}$ for $1 \leq i \leq q$.
It is not hard to see that such a partition exists: $P_1$ included the first $\sqrt{n}$
labelled edges, $P_2$ includes the second $\sqrt{n}$ labelled edges and so on.

The order in which elements of $Var(P_i)$ occur on $P_i$
in fact determines a permutation $SV_i$ of $Var(P_i)$. Any prefix $SV'$ of $SV_i$
clearly corresponds to a prefix of $P_i$ where $SV'$ is the set of variables occurring 
on it. We therefore apply Theorem \ref{mainptv} directly to $P_i$ (rather than to $SV_i$)
and conclude that each $P_i$ has a prefix $P'_i$ such that there is a matching $M'_i$
between $Var(P'_i)$ and the rest of $G$ such that $|M_i| \geq k\log(\sqrt{n})/c_2=k \log n/2c_2$,
where $c_2$ is the constant as in Theorem \ref{mainptv}.

Let $\{u,v\} \in M'_i$. Then one of the ends of the edge, say $u$ belongs to $Var(P'_i)$.
For $v$, there are three possible occurrences: on $P_i \setminus P'_i$,
on $P$ before $P_i$ and on $P$ afer $P_i$. Clearly,  there is $M_i \subseteq M'_i$,
$|M_i| \geq |M'_i| /3$ such that all the `non $P'_i$' ends of the edges of $M_i$ occur in exactly one  one of
these three locations. Let us call the location of non $P'_i$ ends of the edges
of $M_i$ the \emph{popular} location.

If the popular location is $P_i \setminus P'_i$ then let $a_i$ be the last vertex 
of $P_i$. If the popular location is on $P$ before $P_i$ then let $a_i$ be the first
vertex of $P_i$. Finally, if the popular location is on $P$ after $P_i$ then let $a_i$
be the last vertex of $P_i$. Clearly, in any case, $a_i$ separates the ends of each
edge of $M_i$. By Proposition \ref{fixedset}, there is a set $U_i \subseteq V(G)$
including exactly one end of each edge of $M_i$ such that for each source-sink  path $Q$ 
of $Z$ passing through $a_i$, $U_i \subseteq A(Q)$.
Let $U=U_1 \cup \dots U_q$. Clearly, for any source-sink path $Q$ of $Z$ passing through
all of $a_1, \dots, a_q$, $U \subseteq A(Q)$.

It remains to verify that the size of $U$ satisfies the required lower bound.
To this end, let $M=M_1 \cup \dots \cup M_q$. Note that $U$ \emph{covers} $M$
that is each edge of $M$ is incident to at least one vertex of $U$.
Since $G$ of of max-degree $7$, a vertex of $U$ cannot cover more than $7$ edges .
It follows that $|U| \geq |M|/7$. Consequently, it is sufficient to verify that
$|M|=\Omega(k \log n)$.

 To this end, observe that each edge $\{u,v\} \in M$ can belong to at most
two different $M_i$. Indeed, assume that there are distinct $i_1,i_2,i_3$ such that
$\{u,v\} \in M_{i_1} \cap M_{i_2} \cap M_{i_3}$. By definition of
$M_i$ all of $Var(P_{i_1}),Var(P_{i_2}),Var(P_{i_3})$ must intersect with
$\{u,v\}$. A simple pigoenhole pricniple implies that one of $\{u,v\}$
must occur in at least two of  $Var(P_{i_1}),Var(P_{i_2}),Var(P_{i_3})$.
However, this is a contradiction because by definition 
$Var(P_1), \dots, Var(P_q)$ are pairwise disjoint!

Since each edge of $M$ contributes to at most two different $M_i$,
$|M| \geq (\sum_{i=1}^q |M_i|)/2 \geq k \log n/12c_2$, as required.

If $\sqrt{n}$ is not an integer number, we partition $P$ into subpaths
$_1, \dots, P_{r}$ so that $Var(P'_1), \dots, Var(P'_{r-1})$ are of size
$\lceil \sqrt{n} \rceil$, set $q=r-1$ and define $P_i=P'_i$
for $1 \leq i \leq q-1$ and $P_q=P'_{q} \cup P'_{q+1}$.
This way we ensure each $Var(P_i)$ is of size at least $\sqrt{n}$ 
(this is needed for application of Lemma \ref{mainptv})
and $q \leq \Theta(\sqrt{n})$ for . Then we apply the reasoning as above.
$\blacksquare$ 

{\bf Proof of Theorem \ref{mainbound}.}
As in Lemma \ref{bigmatchings}, we assume that $\sqrt{n}$ is integer and then outline
a way to adjust the proof to the general case.  

For each source-sink path $P$ of $Z$, fix vertices $a_1, \dots, a_q$ 
as per Lemma \ref{bigmatchings}, where $q=\sqrt{n}$.
Assume w.l.o.g. that these vertices occur on $P$ in the order listed.
Denote the tuple $(a_1, \dots, a_q)$ by ${\bf a}(P)$ and call it
\emph{the characteristic tuple} of $P$. Recall that $a_i$ is called the
$i$-th \emph{component} of ${\bf a}(P)$.

Let $TP$ be the set of characteristic tuples of all the source-sink 
paths of $Z$. For $1 \leq i \leq q$, let $B_i$ be the set of all $i$-th
components of elements of $TP$ and let $\mu$ be the size of the largest
component. We treat the sets $B_1, \dots, B_q$ are the 
\emph{bottlenecks} of $Z$. Then $\mu$ is the size of the largest bottleneck.
In the rest of the proof we demonstrate that $\mu$ must be large, implying
the theorem. Note that this will tell us that a large bottleneck exist
but will \emph{not} point out to a particular large bottleneck.

For each ${\bf a} \in TP$, let $F_{\bf a}$ be the function whose satisfying
assignments are exactly those of $F$ that are carried by paths passing through
all the components of ${\bf a}$. Observe that for each satisfying assignment $S$
of $F$ there is ${\bf a} \in TP$ such that $S \in F_{\bf a}$. Indeed, let
$P$ be a source-sink path of $Z$ carrying $S$. Then, by definition, $S \in F_{{\bf a}(P)}$.
Therefore $F=\bigcup_{{\bf a} \in TP} F_{{\bf a}}$ and hence 
$|F| \leq \sum_{{\bf a} \in TP} |F_{{\bf a}}|$. Let ${\bf b} \in TP$ be such that
$|F_{{\bf b}}|$ is the largest. Then 
\begin{equation} \label{eq21}
|F| \leq |F_{{\bf b}}|*|TP|
\end{equation}

Let $G$ be the graph such that $\varphi=\varphi(G)$.
Then, according to Theorem \ref{bigmatchings}, there is $U \subseteq V(G)$ of
size at most $k \log n \sqrt{n}/a_1$ for some constant $a_1$ such that 
$F_{{\bf b}} \subseteq F \leftarrow U$. Clearly, $F \leftarrow U \subseteq \varphi(G) \leftarrow U$.
By Theorem \ref{manyvars1}, $|\varphi(G) \leftarrow U| \leq |\varphi|/2^{k \log n \sqrt{n} /b_7*a_1}$.
Therefore, denoting $b_7*a_1$ by $a_2$, we obtain
\begin{equation} \label{eq22}
|F_{{\bf b}}| \leq |\varphi|/(n^{k \log n/a_2})^{\sqrt{n}}
\end{equation}

Each tuple of $TP$ is obtained by taking one element of $B_1$, one element of $B_2$, $\dots$,
one element of $B_q$. Therefore, 

\begin{equation} \label{eq23}
|TP| \leq \prod_{1 \leq i \leq q} |B_i|=\mu^{q}=\mu^{\sqrt{n}}
\end{equation}

Substituting \eqref{eq22} and \eqref{eq23} into \eqref{eq21},
we obtain 
\begin{equation} \label{eq24}
|F| \leq |\varphi|*(\mu/n^{k \log n/a_2})^{\sqrt{n}}
\end{equation}

By our assumption, $|F| \geq |\varphi|/2^{\sqrt{n}}$.
Combining this with \eqref{eq24}, we obtain that 
$|\varphi|*(\mu/n^{k \log n/a_2})^{\sqrt{n}} \geq |\varphi|/2^{\sqrt{n}}$,
hence $(\mu/n^{k \log n/a_2})^{\sqrt{n}} \geq (1/2)^{\sqrt{n}}$, hence
$\mu/n^{k \log n/a_2} \geq 1/2$ from where the desired lower bound for $\mu$
immediately follows.

If $\sqrt{n}$ is not an integer then, by Lemma \ref{bigmatchings}, there is
a unuversal constant $c$ such that for each sufficiently large $n$,
$q \leq c\sqrt{n}$. Replace the equality $\mu^{q}=\mu^{\sqrt{n}}$
in \eqref{eq23} by $\mu^{q} \leq \mu^{c\sqrt{n}}$ 
and update the rest of the reasoning accordingly.
$\blacksquare$
%\bibliographystyle{plain}
%\bibliography{KnowComp}

\appendix
\section{Proof of Theorem \ref{manyvars1}} \label{sec:proportion}
\begin{definition}[{\bf Decision tree}]
Let $F$ be a function that is not constant zero.
Then a \emph{decision tree} $T$ for $F$ is defined as follows.
Suppose that $Var(F)=\{x\}$.
Then the root $rt$ of $T$ is labelled with $x$. If both $\{x\}$ and $\{\neg x\}$ are satisfying
assignments of $F$ then $rt$ has two outgoing edges labelled with $x$ and $\neg x$, respectively
(the leaves are not labelled with variables).
If the only satisfying assignment of $F$ is $\{x\}$ then the only outgoing edge of $rt$ is labelled
with $x$. Finally, if the only satisfying assignment of $rt$ is $\{\neg x\}$ then the only outgoing edge
is labelled with $\neg x$.

Now, assume that $|Var(F)|>1$. Let $x \in Var(F)$.
Label $rt$ with $x$. If both $x$ and $\neg x$ occur in satisfying assignments
of $F$ then $rt$ has two outgoing edges labelled with $x$ and $\neg x$.
Let $u_1$ and $u_2$ be the respective heads of these edges.
Then the subtree of $T$ rooted by $u_1$ is a decision tree for $F|_x$
and the subtree rooted by $u_2$ is a decision tree for $F|_{\neg x}$.

If there is exactly one $\ell \in \{x,\neg x\}$ in the satisfying assignments of $F$ then
there is only one outgoing edge labelled with $\ell$
and the subtree rooted by the head of this edge is a decision tree for $F|_{\ell}$.

For a path $P$ of $T$, we denote (similarly to NROBPs) by $A(P)$ the set of literals
labelling the edges of $P$. 
\end{definition}

%For the rest of this section we define the number of satisfying assignments of $F$
%by $|F|$.

\begin{definition}[{\bf Solutiion counting decision tree}]
A \emph{solution counting decision tree} (SCDT) $T$ for $F$ is a decision
tree for $F$ whose edges are associated with weights defined as follows.
Let $rt$ be the root of $T$ and let $e$ be an outgoing edge of $rt$ labelled
with a literal $\ell$. Then the weight of $e$ is $|F|_{\ell}|/|F|$.

Let $u$ be an internal node of $T$ and let $P$ be the unique $rt-u$ path of $P$
Let $e$ be an outgoing edge of $u$ and let $\ell$ be the literal labelling $e$.
Then the weight of $e$ is $|F|_{A(P) \cup \{\ell\}}|/|F|_{A(P)}|$

The weight of a path of $T$ is the product of weights of its edges.
The weight of an edge $e$ is denoted by $weight(e)$, the weight of a path
$P$ is denoted by $weight(P)$. Finally if ${\bf P}$ is a set of paths then
the weight of ${\bf P}$, denoted by $weight({\bf P})$, is the sum of weights of paths
in ${\bf P}$. 
\end{definition}

\begin{lemma} \label{correctcount}
Let $T$ be a SCDT for a Boolean functon $F$
and let $P$ be a root-leaf path of $T$.
Then $weight(P)=1/|F|$.
\end{lemma}

{\bf Proof.}
By induction on $|Var(F)|$. If $|Var(F)|=1$, this can be verified by a direct inspection.
Assume that $|Var(F)|>1$. Let $P$ be a root-leaf path. Let $e$ be the first edge of $P$ and let $\ell$
be the literal labelling $e$. Let $u$ be the head of $e$ and let $T'$ be the subtree of $T$
rooted by $u$. It is not hard to observe that $T'$ is a SCDT for
$F|_{\ell}$. %probably need to proven as a separate claim or at least stated. 
Let $P'$ be the suffix of $P$ starting at $u$. By the induction assumption,
$weight(P')=1/|F|_{\ell}|$. Now $weight(P)=weight(P')*weight(e)=(1/|F|_{\ell}|)*(|F|_{\ell}|/|F|)=1/|F|$
as required.
$\blacksquare$ 

In light of Lemma \ref{correctcount}, if we need to estimate a proportion of a certain set of satisfying
assignments of a Boolean function, we can calculate the weight of paths carrying these assignments
in the SCDT of this Boolean function.
This is exactly the approach we are taking for proving Theorem \ref{manyvars1}.
From now on, fix $T$ a SCDT for $\varphi(G)$.
Let us introduce additional notation related to $T$.

Let $u$ be a node of $T$. Let $P$ be the unique root-$u$ path of $T$.
We denote $A(P)$ by $A_u$. Let $x \in V(G)$. We denote by $N^u(x)$
the set of neighbours $y$ of $x$ such that $y$ is neither assigned nor forced
to $1$ by $A_u$ (that no neighbour of $x$ occurs negatively in $A_u$).
We let $F^u=\varphi(G)|_{A_u}$.

%\begin{lemma} \label{projsize}
%Let $V \subseteq Var(F)$.
%Let ${\bf S}=Proj(F,Var(F) \setminus F)$.
%Then $|{\bf S}|/|F| \geq 1/2^{|V|}$.
%\end{lemma}

%Each $S \in {\bf S}$ can be extended in $2^{|V|}$ to a set of literals of $Var(F)$.
%Let ${\bf S^*}$ be the set of all such extnesions for the elements of ${\bf S}$.
%Clearly $|{\bf S^*}|=|{\bf S}|*2^{|V|}$ and $F \subseteq {\bf S^*}$.
%$\blacksquare$

%Let $G$ be a graph, let $\varphi_G$ be the monotone CNF with clauses $(x \cup y)$
%corresponding to edges $\{x,y\}$ of $G$.
%Let $T$ be a solutions counting decision tree for $\varphi_G$.

\begin{lemma} \label{largeportion}
Let $x$ be a variable of $V(G)$ labelling $u$.
Assume that $x$ is not forced to $1$ by $A_u$.
Then $|F^u|_{\neg x}|/|F^u| \geq (1/2)^{|N^u(x)|+1}$.
Also, $|F^u|_x|/|F^u| \geq 1/2$.
\end{lemma}

{\bf Proof.}
Note first that since $A_u$ is an assigned labelling a path of $T$, it has 
a satisfying extension. That is $|F_u|>0$, hence the left-hand sides of the
desired inequalities are defined. 

\begin{claim}\label{projsize}
Let $F$ be a Boolean function
Let $V \subseteq Var(F)$.
Let ${\bf S}=Proj(F,Var(F) \setminus V)$.
Then $|{\bf S}|/|F| \geq 1/2^{|V|}$.
\end{claim}

{\bf Proof.}
Each $S \in {\bf S}$ can be extended in $2^{|V|}$ to a set of literals of $Var(F)$.
Let ${\bf S^*}$ be the set of all such extensions for the elements of ${\bf S}$.
Clearly $|{\bf S^*}|=|{\bf S}|*2^{|V|}$ and $F \subseteq {\bf S^*}$.
$\square$

Let $V^u=Var(F^u) \setminus (N^u(x) \cup \{x\})$.
In other words, $V^u$ is obtained from the set of variables not assigned by $A_u$
by removal $x$ and those neighbours of $x$ that are forced to $1$ by $A^u$.

\begin{claim} \label{portion1}
$Proj(F^u,V^u) \subseteq Proj(F^u|_{\neg x},V^u)$. 
\end{claim}

{\bf Proof.}
Let $S \in Proj(F^u,V^u)$. Let $A^*$ be the set of literals over 
$N^u(x) \cup \{x\}$ assigning $x$ negatively and the rest of the variables
positively. Note that $Var(S)$, $Var(A_u)$ and $Var(A^*)$
partition $V(G)$. 

We claim that $S^*=S \cup A_u \cup A^*$ is a satisfying assignment 
of $\varphi(G)$. To prove this let us show that each clause $(y \cup z)$
of $\varphi(G)$ is satisfied by $S^*$.

Note that by definition of $S$, $A_u \cup S$ has a satisfiable extension
to the rest of the variables of $V(G)$. Therefore, the claim holds if both
$y$ and $z$ belong to $Var(S \cup A_u)$. The same is true with $Var(A^*)$
simply because $A^*$ contains negative occurrence of exactly one variable.
It remains to assume that one variable, say, $y$ is contained in $Var(A^*)$
and the other, $z$, is contained in $Var(S \cup A_u)$.
Then $y=x$ because otherwise $y$ is positively assigned and the clause is satisfied.
Then $z$ is a neighbour of $x$. If $z \in Var(A_u)$ then $z$ occurs positively
in $A_u$ because otherwise $x$ is forced to $1$ in contradiction to our assumption.
If $z \in Var(S)$ then, as $z \notin N^u(x)$, we conclude that $z$ is forced to 
$1$ by $A_u$ and hence also positively assigned.
Now, since $S^*$ contains $\neg x$, $S^* \setminus (A_u \cup \{\neg x\})$
is a satisfying assignment of $F^u|_{\neg x}$ and hence $S$ belongs to
the projection of $F^u|_{\neg x}$ to $V^u$.
$\square$

It follows from the Claim \ref{portion1} that 
$|Proj(F^u,V^u)| \leq |Proj(F^u|_{\neg x},V^u)|$.
As the size of a set of literals cannot be smaller than
the size of its projection, it also follows that 
$|Proj(F^u,V^u)| \leq |F^u_{\neg x}|$.

As $|Proj(F^u,V^u)|/|F^u| \geq (1/2)^{|N^u(x)|+1}$
according to Claim \ref{projsize}, the second statement of the 
lemma follows. The first statement can be established by the same
approach taking $V^u=Var(F^u) \setminus \{x\}$.
$\blacksquare$

\begin{lemma} \label{treeweights}
With the data as in Lemma \ref{largeportion}, assume that $x$ is not forced by 
$A_u$. Then the wieght of the outgoiing edge of $u$ labelled with $\neg x$
ibetween $(1/2)^{|N^u(x)|+1}$ and $1/2$ and the weight of the outgoing
edge of $u$ labelled with $x$ is between $1/2$ and $1-(1/2)^{|N^u(x)|+1}$.
\end{lemma}

{\bf Proof.}
Immediately from Lemma \ref{largeportion}.
Lower bounds are given directly. The upper bound is $1$ minus the lower bound 
for the opposite literal.
$\blacksquare$

For $d \geq 0$, let $c_d=1-2^{-(2d+1)}$.
%Let $G$ be a graph with max-degree $d$ 
%and let $T$ be a solutions counting decision tree for $\varphi_G$.
%$G$ and $T$ will be fixed for the reasoning below.
For $S \subseteq V(G)$, we denote $\prod_{x \in S} c_{|N^u(x)|}$ by $\alpha^u(S)$.

Let $u$ be a vertex of $T$,
$S \subseteq V(G)$.
We denote by ${\bf P}^u(S)$ the set of paths $P$ of $T$ from $u$ to 
a leaf such that $S \subseteq A(P)$.

\begin{theorem} \label{maintree}
Let $u$ be a vertex of $T$ 
labelled by a variable $x$
and let $S \subseteq V(G)$ such that 
none of elements of $S$ are neighbours or having common neighbours and
none of elements
of $S$ is forced to $1$ by $A_u$.
Then $weight({\bf P}^u(S)) \leq \alpha^u(S)$.
\end{theorem}

The proof of Theorem \ref{maintree} is quite tedious.
Therefore, we first show that how Theorem \ref{manyvars1}
follows from it. 

{\bf Proof of Theorem \ref{manyvars1}.}
Recall that $d$ is the max-degree of $G$.
It is not hard to see that there is a subset $S$ of $U$ of size
at least $|U|/(d+1)$ such that no two elements of $S$ are adjacent.

Clearly, for each $x \in V(G)$, $|N^u(x)| \leq d$. 
Hence, it follows from Theorem \ref{maintree} that for each $u \in V(T)$,
$weight({\bf P}^u(S)) \leq c_d^{|S|}$.
In particular, $weight({\bf P}^{rt}(S)) \leq c_d^{|S|}$
where $rt$ is the root of $T$.
Note that $\{A(P)|P \in {\bf P}^{rt}(S)\}=\varphi(G)|_{S}$.
It follows from Lemma \ref{correctcount} that 
$weight({\bf P}^{rt}(S))=|\varphi(G)|_{S}|/|\varphi(G)|$.

Note that $|\varphi(G) \leftarrow S|=\varphi(G)|_S$.
Then we conclude that
$|\varphi(G) \leftarrow S|/|\varphi(G)| \leq c_d^{|S|} \leq c_d^{|U|/(d+1)} \leq
(c_d^{1/(d+1)})^{|U|}$. 
Then $|\varphi(G) \leftarrow S| \leq |\varphi(G)|/2^{|U|/b_d}$ where
$2^{b_d}=(1/c_d)^{1/(d+1)}$.
$\blacksquare$

\begin{lemma} \label{smalleralpha}
Let $u$ be a vertex of $T$, let $v$ be its child.
Let $S \subseteq V(G)$. Then $\alpha^v(S) \leq \alpha^u(S)$.
\end{lemma}

{\bf Proof.}
We only need to verify that for each $x \in V(G)$,
$c_{|N^v(x)|} \leq c_{|N^u(x)|}$. Indeed, any neighbour of $x$
that is assigned or forced to $1$ by $A_u$ retains this status
regarding $A_v$. Hence $N^v(x) \subseteq N^u(x)$ implying that 
$|N^v(x)| \leq |N^u(x)|$. The lemma now follows immediately
from an easy to observe fact that if $d'<d$ then $c_{d'} \leq c_d$.
$\blacksquare$

\begin{lemma} \label{directweight}
Let $u$ be a vertex of $T$ labelled by $x$.
Let $S \subseteq V(G)$ such that $x \in S$.
Let $v$ be the positive outneighbour of $u$ and let $p$ be the weight of $(u,v)$.
Then $p*\alpha^v(S \setminus \{x\}) \leq \alpha^u(S)$.
\end{lemma}

{\bf Proof.}
By definition, $\alpha^u(S)=c_{|N^u(x)|}*\alpha^u(S \setminus \{x\})$.
By Lemma \ref{smalleralpha}, $\alpha^v(S \setminus \{x\}) \leq \alpha^u(S \setminus \{x\})$.
Hence, it remains to verify that $p \leq c_{|N^u(x)|}$. 
By Lemma \ref{treeweights}, $p \leq 1-(1/2)^{|N^u(x)|+1} \leq c_{|N^u(x)|}$.
$\blacksquare$

\begin{lemma} \label{neighbweight}
Let $u$ be a vertex of $T$ labelled by $x$.
Let $S \subseteq V(G)$ contains a neighbour $y$ of $x$.
Assume that $x$ is not forced to $1$ by $A_u$ and let $vp$ and $vn$
be respective positive and negative out-neighbours of $u$.
Let $p$ be the weight of $(u,vp)$.
Then $p*\alpha^{vp}(S)+(1-p)*\alpha^{vn}(S \setminus \{y\}) \leq \alpha^u(S)$.
\end{lemma}

{\bf Proof.}
By definition $\alpha^{vp}(S)=c_{|N^{vp}(y)|}*\alpha^{vp}(S \setminus \{y\})$
and $\alpha^{u}(S)=c_{|N^{u}(y)|}*\alpha^{u}(S \setminus \{y\})$.
As both $\alpha^{vp}(S \setminus \{y\})$ and $\alpha^{vn}(S \setminus \{y\})$
do not exceed $\alpha^{u}(S \setminus \{y\})$ by Lemma \ref{smalleralpha},
it follows that it is sufficient to verify that 
$p*c_{|N^{vp}(y)|}+(1-p) \leq c_{|N^{u}(y)|}$.

To this end, note that that by our assumption, $x \in N^u(y)$ while
$x \notin N^{vp}(y)$. As $N^{vp}(y) \subset N^u(y)$,
$|N^{vp}(y)| \leq |N^u(y)|-1$. As $c_d$ is a non-decreasing function in $d$,
we conclude that $c_{|N^{vp}(y)|} \leq c_{|N^u(y)|-1}$.
Thus $p*c_{|N^{vp}(y)|}+(1-p) \leq p*c_{|N^u(y)|-1}+(1-p)=
p*(1-(1/2)^{2|N^u(y)|-1})+1-p=1-p*(1/2)^{2|N^u(y)|-1})$.
The rightmost part of the above derivation grows with decrease of $p$.
As $p \geq 1/2$ by Lemma \ref{treeweights}, 
$1-p*(1/2)^{2|N^u(y)|-1}) \leq 1-(1/2)^{2|N^u(y)|})<c_{|N^u(x)|}$.
$\blacksquare$

Let $v$ be a vertex of $T$ and let ${\bf P}$ be a set of paths, all starting 
from $v$. Let $(u,v)$ be an edge of $T$.
Then $(u,v)+{\bf P}$ denotes the set of paths 
$\{(u,v)+P|P \in {\bf P}\}$. 
Clearly, $weight((u,v)+{\bf P})=weight(u,v)*weight({\bf P})$.

\begin{lemma} \label{redundvert}
Let $u$ be a non-leaf vertex of $T$.
Assume that $u$ is labelled with a variable $x \notin S$.
Let $vp$ be the positive out-neighbour of $u$.
If $vp$ is the only out-neigbhour of $u$ then 
${\bf P}^u(S)=(u,vp)+{\bf P}^{vp}(S)$.
Otherwise, let $vn$ be the negative out-neighbour of $u$.
In this case,
${\bf P}^u(S)=[(u,vp)+{\bf P}^{vp}(S)] \cup [(u,vn)+{\bf P}^{vn}(S)]$
\end{lemma}

{\bf Proof.}
Let $P \in {\bf P}^u(S)$ and let $P'$ be the suffix of $P$ starting from
an outneighbour of $u$. Then $S \subseteq A(P')$ because otherwise, since
$x \notin S$, $S$ is not a subset of $A(P)$ either.

Conversely, if $S\subseteq A(P')$ where $P'$ is a path starting from
an out-neighbour of $u$. Then, clearly, $S \subseteq A((u,v)+P')$
and hence $(u,v)+P' \in {\bf P}^u(S)$.
$\blacksquare$

\begin{lemma} \label{insidevert}
Let $u$ be a non-leaf vertex of $T$.
Assume that $u$ is labelled with a variable $x \in S$.
Let $vp$ be the positive out-neighbour of $u$.
${\bf P}^u(S)=(u,vp)+{\bf P}^{vp}(S \setminus \{x\})$.
\end{lemma}

{\bf Proof.}
Assume that 
$P \in (u,v)+{\bf P}^{vp}(S \setminus \{x\})$.
Then, by assumption, $(u,vp)$ contributes $x$ to $A(P)$ and the rest of the edges 
contribute $S \setminus \{x\}$, hence $S \subseteq A(P)$.

Conversely, assume that $P \in {\bf P}^u(S)$ and let $v$ be the
immediate successor of $u$ on $P$. If $v \neq vp$ then 
$v$ is the negative out-neighbour of $u$. Hence $\neg x \in A(P)$, that
is $x \notin A(P)$ (due to read-onceness) and hence, in particular,
$S \nsubseteq A(P)$, a contradiction. It remains to assume that $v=vp$.
Then, as $(u,v)$ contributes $x$ to $A(P)$, it does not contribute to 
$S \setminus \{x\}$ and hence this must be contributed by the prefix of
$P$ starting at $vp$.
$\blacksquare$

\begin{lemma} \label{neighbvert}
Let $u$ be a non-leaf vertex of $T$.
Assume that $u$ is labelled with a variable $x \notin S$ 
and a neighbour of $y \in S$ such that $y$ does not
occur in $A_u$.
Assume further that $u$ has two outneighbours.
Denote the positve and negative outneighbours of $u$ by $vp$ and
$vn$, respectively. 
Then 
${\bf P}^u(S)=[(u,vp)+{\bf P}^{vp}(S)] \cup [(u,vn)+{\bf P}^{vn}(S \setminus \{y\})]$.
\end{lemma}

{\bf Proof.}
As 
$[(u,vp)+{\bf P}^{vp}(S)] \cup [(u,vn)+{\bf P}^{vn}(S)] \subseteq 
[(u,vp)+{\bf P}^{vp}(S)] \cup [(u,vn)+{\bf P}^{vn}(S \setminus \{y\})]$,
${\bf P}^u(S) \subseteq [(u,vp)+{\bf P}^{vp}(S)] \cup [(u,vn)+{\bf P}^{vn}(S \setminus \{y\})]$
by Lemma \ref{redundvert}.

Assume now that $P \in [(u,vp)+{\bf P}^{vp}(S)] \cup [(u,vn)+{\bf P}^{vn}(S \setminus \{y\})]$.
If $P \in  [(u,vp)+{\bf P}^{vp}(S)]$ then apply again Lemma \ref{redundvert}.
So, we assume that the immediate successor of $u$ in $P$ is $vn$.
Let $P'$ be the suffix of $P$ starting at $vn$. We show that $S \subseteq A(P)$.
By assumption, it is enough to show that $y \in A(P)$. By assumption $y$ has not been assigned
by $A_{vn}$ and is forced to $1$.
$\blacksquare$

{\bf Proof of Theorem \ref{maintree}.}
Note that if at least one of the elements of 
$S$ is assigned by $A_u$ then ${\bf P}^u(S)=\emptyset$.
Hence, in this case $weight({\bf P}^u(S))=0$ and the theorem holds 
as $\alpha^u(S)$ is non-negative by definition. So, we can assume that
no element of $S$ is assigned by $A_u$.

Assume first that $u$ is a leaf. In light of the previous paragraph,
$S=\emptyset$. Hence, ${\bf P}^u(S)=\{u\}$ and 
$weight({\bf P}^u(S))=1$. On the other hand, $\alpha^u(\emptyset)=1$
as well. Hence the theorem holds.

Assume now that $u$ is not a leaf and the theorem holds
for all the descendants of $u$.
Consider first the case where $u$ has exactly one outneighbour.
Then it is a positve outneighbour, denote it $vp$.
Let $x$ be the variable labelling $u$. It follows that $x$
is forced to $1$ by $A_u$, hence $x \notin S$.

It follows from Lemma \ref{redundvert} that
${\bf P}^u(S)=(u,vp)+{\bf P}^{vp}(S)$.
Hence 
$weight({\bf P}^u(S))=weight((u,vp))*weight({\bf P}^{vp}(S))$.
As $vp$ is the only out-neighbour of $u$, the weight
of $(u,vp)$ is $1$.
Hence $weight({\bf P}^u(S))=weight({\bf P}^{vp}(S))$.
Note that no element of $S$ is forced to $1$ by $A_{vp}$
as $A_{vp}=A_{u} \cup \{x\}$, this is true regarding
$A_u$ by assumption and regarding $x$ due to it being a positive 
literal. Hence, we may apply the induction assumption,
according to which $weight({\bf P}^{vp}(S)) \leq \alpha^{vp}(S)$.
As $\alpha^{vp}(S) \leq \alpha^u(S)$ by Lemma \ref{smalleralpha},
the theorem holds in this case.

It remains to assume that $u$ has two out-neighbours.
Denote the positive and negative ones by $vp$ and $vn$, respectively.
Let $p$ be the weight of $(u,vp)$ (and hence the weight of $(u,vn)$
is $1-p$).

Assume first that $x \notin S$ and $x$ is not a neighbour of any element of $S$.
By Lemma \ref{redundvert},
${\bf P}^u(S)=[(u,vp)+{\bf P}^{vp}(S)] \cup [(u,vn)+{\bf P}^{vn}(S)]$.
As $[(u,vp)+{\bf P}^{vp}(S)]$ is disjoint with $[(u,vn)+{\bf P}^{vn}(S)]$,
$weight({\bf P}^u(S))=weight((u,vp)+{\bf P}^{vp}(S))+ weight((u,vn)+{\bf P}^{vn}(S))=
p*weight({\bf P}^{vp}(S))+(1-p)*weight({\bf P}^{vn}(S))$.
As $x$ is not a neighbour of any element of $S$, none of its literals forces any element of $S$
into $1$. Hence, arguing as in the previous case, the induction assumption can be 
applied to both $weight({\bf P}^{vp}(S)$ and $weight({\bf P}^{vn}(S))$.
Hence 
$weight({\bf P}^u(S)) \leq p*\alpha^{vp}(S)+(1-p)*\alpha^{vn}(S) \leq p*\alpha^u(S)+(1-p)*\alpha^u(S)=\alpha^u(S)$,
the last inequality follows from Lemma \ref{smalleralpha}.

Assume now that $x$ is a neighbour of some element $y$ of $S$.
By our assumption about $S$, $x \notin S$ and $y$ is the only neighbour of $x$ in $S$.
As $y$ is not assigned by $A_u$ by our assumption, it follows from Lemma \ref{neighbvert}
that
${\bf P}^u(S)=[(u,vp)+{\bf P}^{vp}(S)] \cup [(u,vn)+{\bf P}^{vn}(S \setminus \{y\})]$.
Arguing as in the previous case, we conclude that
$weight({\bf P}^u(S))=p*weight({\bf P}^{vp}(S))+(1-p)*weight({\bf P}^{vn}(S \setminus \{y\}))$.
As $x$ is a positive literal it does not force to $1$ any element of $S$.
Also, since $x$ does not have neighbours in $S \setminus \{y\}$, $\neg x$ does not force
any fo elements of $S \setminus \{y\}$ to $1$.
It follows that the induction assumption can be applied to 
$weight({\bf P}^{vp}(S))$ and $weight({\bf P}^{vn}(S \setminus \{y\}))$.
Thus we obtain 
$weight({\bf P}^u(S)) \leq p*\alpha^{vp}(S)+(1-p)*\alpha^{vn}(S \setminus \{y\}) \leq \alpha^u(S)$,
the last inequality follows from Lemma \ref{neighbweight}.

It remains to assume that $x \in S$. 
By Lemma \ref{insidevert}
${\bf P}^u(S)=(u,vp)+{\bf P}^{vp}(S \setminus \{x\})$,
and hence
$weight({\bf P}^u(S))=p*weight({\bf P}^{vp}(S \setminus \{x\})$.
Being a positive literal, $x$ does not force any variable of $S \setminus \{x\}$
into $1$, hence the induction assumption can be applied.
It follows that
$weight({\bf P}^u(S)) \leq p*\alpha^{vp}(S \setminus \{x\}) \leq alpha^u(S)$,
the last inequality follows from Lemma \ref{directweight}.
$\blacksquare$

\end{document}